\documentclass[aps,prl,twocolumn,showkeys,showpacs]{revtex4}
\usepackage{epsfig}

\begin{document}

\title{Minimal coupling in oscillator models of quantum dissipation}

\author{H.~Kohler$^1$, F.~Sols$^2$ }
\email{hkohler@icmm.csic.es}
\email{f.sols@fis.ucm.es}
\affiliation{$^1$ Instituto de Ciencia de Materiales de Madrid, CSIC,
Sor Juana de la Cruz 3, Cantoblanco, 28049 Madrid, Spain\\
$^2$ Departamento de F\'{\i}sica de Materiales, Universidad
Complutense de Madrid, E-28040 Madrid, Spain.
}
\date{\today}

\begin{abstract}
The dissipative harmonic oscillator has two representations. In the first representation the central oscillator couples with its position to an oscillator bath. In the second one it couples with its momentum to the bath. Both representations are related by a unitary transformation. If the central oscillator couples  with its position and momentum  to two independent baths, no such unitary transformation exists. We discuss two possible models of this type and their physical relevance.
\end{abstract}


\maketitle

The Hamiltonian of the harmonic oscillator with mass $m$ and frequency $\omega_0$ is characteristically symmetric in position and momentum.
On the other hand, the dissipative harmonic oscillator serves as an example of quantum dissipation induced by a gauge field. Let us assume local $U(1)$ gauge invariance, requiring expectation values to be invariant under the gauge transformation of the wave function $\psi(q)\to \exp[i\Lambda(q)]\psi(q)$. This is achieved by introducing a gauge field $A(q)$ which transforms as $A(q)$ $\to$ $A(q)-\partial_q \Lambda(q)$. The kinematic momentum
\begin{equation}
 p \to p+ A(q) \
\end{equation}
is gauge invariant. The gauge field $A$ might be decomposed into normal modes and be quantized (hereafter $\hbar=1$)
\begin{equation}
\hat{A} \ =\ i \sum_{k}\frac{\lambda_k}{\omega_k}
         \left(\hat{a}_{k} e^{ikq} - \hat{a}^\dagger_{k}e^{-ikq}\right)\ ,
\end{equation}
where $\hat{a}_{k}$ and $\hat{a}^\dagger_{k}$ are bosonic creation and annihilation operators of modes with frequency $\omega_k$. The parameters $\lambda_k$ characterize the strength of the coupling of the gauge field to matter. The resulting Hamiltonian reads
\begin{eqnarray}
\label{Hmom}
\hat{H} &= &\frac{1}{2m}\left[\hat{p}+ \hat{A}(q)\right]^2 + \frac{m\omega_0^2}{2}{\hat{q}}^2 + \sum_k \omega_k \hat{a}_k^\dagger \hat{a}_k\label{com1} \ ,
\end{eqnarray}
with the gauge field minimally coupled to the canonical momentum. In the form (\ref{Hmom}) the canonical momentum $\hat{p}$ and the gauge potential $\hat{A}$ appear as dynamical variables and both quantities are gauge dependent.

 We consider in the following the long wavelength approximation $k\to 0$. The space dependence disappears $e^{ikq}\simeq 1$ and
 \begin{equation}
 \label{Aquant}
 \hat{A}\ = \ \sum_k i\frac{\lambda_k}{\omega_k}(\hat{a}_k-\hat{a}_k^\dagger).
 \end{equation}
 A unitary transformation $\hat{U}$ $=$ $\exp\left(i\hat{A}\,\hat{q}\right)$ can be applied on $\hat{H}$. This is the so-called polaron or  G{\"o}ppert--Mayer transformation \cite{coh89}. It brings the Hamiltonian into the standard form of the dissipative harmonic oscillator
\begin{eqnarray}
\hat{H}^{\prime} &= &\frac{\hat{p}^2}{2m} + \frac{m \omega_0^2}{2} \hat{q}^2 + \sum_k\omega_k\left|\hat{a}_k+\frac{\lambda_k}{\omega_k} \hat{q}\right|^2 \ . \label{com2a}
\end{eqnarray}
This Hamiltonian was studied first in \cite{ulle66} and later by many others, see \cite{grab88,wei99} and references therein \cite{comment1,sanc94}. For an arbitrary operator $\hat{O}$ we use the short--hand notation $|\hat{O}|^2$ $=$ $\hat{O}^\dagger \hat{O}$. Now $p$ is the kinematic (gauge invariant) momentum, which in this particular gauge coincides with the (generally gauge-dependent) canonical momentum, and the particle couples via its position to the modes of the gauge field. Thus a specific implementation of quantum dissipation can be derived from a $U(1)$ gauge invariance principle. However, in the presence of the gauge field, the $q$--$p$ symmetry under exchange of position and momentum is broken, the reason being that we assumed gauge invariance which is local in position but not in momentum.

Some authors \cite{leg84,Cuccoli2001} considered as well a Hamiltonian similar to (\ref{com2a}), where the oscillator bath couples linearly to the momentum
\begin{eqnarray}
\hat{H}^{\prime\prime} &= &\frac{\hat{p}^2}{2m} + \frac{m \omega_0^2}{2} \hat{q}^2 + \sum_k\omega_k\left|\hat{a}_k+\frac{\lambda_k}{\omega_k} \hat{p}\right|^2 \ . \label{com2aa}
\end{eqnarray}
Albeit their formal similarity the physics described by $\hat{H}^{\prime\prime}$ and $\hat{H}^{\prime}$ is quite different. We emphasize that $\hat{H}^{\prime\prime}$ is inequivalent to $\hat{H}$ (in the sense of not being related by a unitary transformation), even with $\hat{A}$ as in Eq.~(\ref{Aquant}) although in both Hamiltonians the oscillator bath couples to the momentum. For the coupling to an oscillator bath as implemented in $\hat{H}^{\prime\prime}$ the term ``anomalous dissipation" was coined \cite{leg84}.

Recently a certain class of dissipative quantum systems has aroused interest, where the system couples to two heat baths through non-commuting variables \cite{cas03,nov05,koh05,koh06b,rao07,cuc10,rao08}. In Refs.~\cite{cas03,nov05,rao08} a spin was considered which couples to two independent oscillator baths. In Refs.~[\onlinecite{koh05,koh06b,cuc10}] the dissipative harmonic oscillator with two competing heat baths was investigated. The model was described by the Hamiltonian
\begin{eqnarray}
\hat{H}_{\rm I} &= &\frac{\omega _{q}}{2}\hat{q}^{2}+\sum_{k}\omega _{k}\left\vert \hat{a}_k+\frac{%
\lambda_{k}}{\omega_{k}}\hat{q}\right\vert ^{2}+\nonumber\\
&&\qquad\quad\frac{\omega _{p}}{2}%
\hat{p}^{2}+\sum_{l}\nu_{l}\left\vert \hat{b}_l+\frac{\mu_{l}}{\nu_{l}}%
\hat{p}\right\vert ^{2}\ , \label{com3}
\end{eqnarray}
which combines both the conventional coupling of $\hat{H}^{\prime}$ and the anomalous coupling of $\hat{H}^{\prime\prime}$. Such a Hamiltonian can be found in real physical situations \cite{koh04,cuc10}. Here $\hat{a}_k^\dagger$ and $\hat{b}_k^\dagger$ are bosonic creation operators of two independent heat baths (bath A and bath B). The frequency of the central oscillator is given by $\omega_0= \sqrt{\omega_p\omega_q}$. In this model the $q$--$p$ symmetry under exchange of position and momentum is reestablished, when $\omega_p$ $=$ $\omega_q$ and the coupling coefficients are identical.

It is tempting to motivate the model (\ref{com3}) by a gauge argument as outlined above for the standard dissipative harmonic oscillator and leading to the equivalence of (\ref{Hmom}) and (\ref{com2a}). A second ``gauge field" $\hat{B}$ is introduced which couples to the position. The most symmetric and most canonical form to introduce this field is via the Hamiltonian
 \begin{eqnarray}
\hat{H}_{\rm II}=\frac{\omega_p}{2}\left(\hat{p}+\hat{A}\right)^2+
            \frac{\omega_q}{2}\left(\hat{q}+\hat{B}\right)^2+\nonumber\\
            \sum_{k}\omega _{k}\hat{a}_k^{\dagger }\hat{a}_k +
             \sum_{l}\nu_l\hat{b}_l^{\dagger
}\hat{b}_l \ ,  \label{mod1}
\end{eqnarray}%
where in the long wavelength approximation $\hat{B}$ is given by $\hat{B}$ $=$ $i\sum_{l}\frac{\mu_{l}}{\nu_l}(\hat{b}_l-\hat{b}_l^{\dagger
})$. However, the Hamiltonians $\hat{H}_{\rm I}$ and $\hat{H}_{\rm II}$ are not equivalent in the sense that they cannot be transformed into each other through a canonical transformation. As a consequence, the claim made in Refs.~\cite{koh05,koh06b,koh04} that $H_{\rm I}$ and $H_{\rm II}$ are unitarily equivalent is inaccurate. The analysis there performed corresponds to the $H_{\rm I}$ model, for which  a physical realization was found \cite{koh04}. The subsequent analysis of the Hamiltonian~(\ref{com3}) is correct and the results in Refs.~\onlinecite{koh04,koh05,koh06b} remain unaffected.

One might naively argue that the Hamiltonian $\hat{H}_{\rm I}$ can be transformed into $\hat{H}_{\rm II}$ by the coordinate transformation $\hat{q}_{\rm II} = \hat{q}_{\rm I}-\hat{B}_{\rm I}$, $\hat{p}_{\rm II} = \hat{p}_{\rm I}-\hat{A}_{\rm I}$, $\hat{a}_{k,{\rm II}} = \hat{a}_{k,{\rm I}}+\lambda_k \hat{q}_{\rm I}/\omega_k$, and $\hat{b}_{l,{\rm II}} = \hat{b}_{l,{\rm I}}+\mu_l \hat{q}_{\rm I}/\nu_l$, where the lower indices I and II refer to the corresponding Hamiltonians. But, since $\hat{b}_{l,{\rm II}}$ and $\hat{a}_{k,{\rm II}}$, as obtained by this transformation, do not commute, this transformation is not canonical.

To see this even more clearly one can diagonalize both Hamiltonians and verify that their spectra are different.  We define the characteristic polynomial $\chi_{\rm I,II}^{-1}(\omega) \ = \ \det\left(\omega-{\cal H}^{\rm (I,II)}\right) $. Here ${\cal H}^{\rm (I,II)}$ is the matrix which governs the time evolution of the vector whose components are the creation and annihilation operators appearing in the Hamiltonian $\hat{H}_{\rm I}$ (defined as model I in the following) respectively in  $\hat{H}_{\rm II}$ (defined as model II in the following).
The frequencies of the normal modes are given by the positive roots of the characteristic equation $\chi_{\rm I,II}^{-1}(\omega) \ =\  0$. Diagonalising ${\cal H}^{\rm (II)}$ yields

\begin{equation}
\chi_{\rm II} ^{-1}(\omega )=\omega _{0}^{2}-\left[\omega + \frac{\omega _{q}}{\omega}\widetilde{J}_{p}(\omega )\right]
                                    \left[\omega +\frac{\omega _{p}}{\omega}\widetilde{J}_{q}(\omega )\right]\label{sus2} \ ,
\end{equation}
where we have defined the spectral functions
\begin{eqnarray}
 J_q (\omega) & \equiv & 2\sum_k |\lambda_k|^2\delta(\omega-\omega_k) \, \nonumber\\
 J_p(\omega) & \equiv & 2\sum_l |\mu_l|^2\delta(\omega-\nu_l)
\end{eqnarray}
and, as in \cite{koh06b}, the Riemann-Stieltjes integral transform
\begin{equation}
\widetilde{f}(\omega )\ \ \equiv\ \omega ^{2}\mathcal{P}\int_{0}^{\infty }\frac{%
f(\omega ^{\prime })}{\omega ^{\prime }\left( {\omega ^{\prime }}^{2}-\omega
^{2}\right) }d\omega ^{\prime }-i\mathrm{sgn\,}(\omega )\frac{\pi }{2}%
f(|\omega |)~.  \label{tildetransformed}
\end{equation}%
has been introduced. On the other hand, for model I it was found (see \cite{koh06b})
\begin{equation}
\chi_{\rm I}^{-1}(\omega )=\left[\omega _{q}-\widetilde{J}_{q}(\omega )\right]
                                    \left[\omega_p -\widetilde{J}_{p}(\omega )\right]-\omega^2 \ . \label{sus2a}
\end{equation}
Comparison of Eqs.~(\ref{sus2}) and (\ref{sus2a}) shows that the characteristic polynomials and therefore the spectra of both models are in general different.
Importantly, they become identical if  $J_q(\omega)$ or $J_p(\omega)$ vanishes.



The question of which of the Hamiltonians $\hat{H}_{\rm I}$ or $\hat{H}_{\rm II}$ is physically relevant has to be answered in favor of the Hamiltonian $\hat{H}_{\rm I}$ as given in Eq.~(\ref{com3}). As already mentioned above, one example for a physical realisation of $\hat{H}_{\rm I}$ was found in \cite{koh04}: In a Josephson junction the system which is represented by the particle number difference and the phase difference across the junction couples to two independent environments given by the bosonic exitations of quasiparticle tunnelling across the junction and by the electromagnetic vacuum modes. It was shown that the corresponding Hamiltonian has the form of $\hat{H}_{\rm I}$.

Another example was given recently by Cuccoli and coworkers for spin Hamiltonians \cite{cuc10,cuc08}. Spin operators can often be treated as ordinary canonical variables. Then an environmental coupling typically involves both the
coordinates and the momenta in a symmetric fashion, with no privileged role. This has been shown in Ref.~\cite{cuc08} for the case of the easy-axis XXZ magnet.

Although so far it seems to have no direct physical application, it is still interesting to analyze model II as  given in Eq.~(\ref{mod1}).  In this brief report we therefore calculate the equilibrium correlation function of $\hat{Q}\equiv \hat{q}+ \hat{B}$ in  model II and compare it with the equilibrium correlation function of $\hat{q}$ in model I. Our goal is to see if and to what extent the physics of both models is different.

The Heisenberg equations for the bath modes can be solved. Plugging the results into the Heisenberg equations for
$\hat{Q}$ and $\hat{P}\equiv \hat{p}+\hat{A}$ yields
\begin{eqnarray}
\dot{\hat{Q}\,}(t) &=&\omega _{p}\hat{P}(t)-\omega_q\int\limits_{-\infty}^{t}ds\,K_{q}(t-s)\hat{Q}(s)+\hat{F}_{q}(t)\nonumber\\
\dot{\hat{P}}(t) &=&-\omega _{q}\hat{Q}(t)-\omega_p\int\limits_{-\infty}^{t}ds\,K_{p}(t-s)\hat{P}(s) +\hat{F}_{p}(t).\nonumber\\
\label{equations of motion}
\end{eqnarray}%
The response kernel is defined as%
\begin{equation}
K_{n}(t)\equiv \int_{0}^{\infty }\frac{J_{n}(\omega )}{\omega}
                     \cos (\omega t)d \omega \ ,\ n=q,p\ .
\label{EM5}
\end{equation}
We defined the force operator
\begin{equation}
\hat{F}_{q}(t)=\sum \lambda _{k}\hat{a}_k\exp (-i\omega
_{k}t)+\mathrm{H.c} \ ,
\end{equation}
with $\hat{F}_{p}(t)$ defined accordingly. In Fourier
space the Heisenberg equations (\ref{equations of motion}) read%
\begin{eqnarray}
\label{eqmoFour}
\omega_q \hat{Q}+\left[i\omega- \frac{\omega_p}{i\omega}\widetilde{J}_{q}(\omega)\right] \hat{P}& = & \hat{F}_{q}(\omega )  \nonumber \\
\left[i\omega- \frac{\omega_q}{i\omega}\widetilde{J}_{p}(\omega)\right] \hat{Q}-\omega_p \hat{P}& = & \hat{F}_{p}(\omega) \ .
\end{eqnarray}%
The equilibrium correlation function  $C_{qq}^{\rm (II)}(t,\beta)%
\equiv \frac{1}{2}\langle \{ \hat{Q}(t),\hat{Q}(0)\}\rangle_\beta $  is
obtained from equation (\ref{eqmoFour}) after Fourier transformation, namely,
\begin{eqnarray}
C_{qq}^{\rm (II)}(t,\beta) &=&\frac{1}{\pi }\int_{0}^{\infty }d\omega |\chi_{\rm II}(\omega )|^{2}\cos
(\omega t)\coth (\beta \omega /2)  \nonumber\\
&& \left[\gamma_p\omega_0^2(1+\gamma_q\gamma_p)\omega + \gamma_q\omega^3\right] \ . \label{symmetrized autocorrelationfkt}
\end{eqnarray}

In the following we set $\omega_p$ $=$ $\omega_q$ $=$ $\omega_0$. Moreover we focus on the case of Ohmic damping and set  $\widetilde{J}_{q}(\omega)$ $=$ $i\gamma_q \omega$ and $\widetilde{J}_{p}(\omega)$ $=$ $i\gamma_p \omega$, where $\gamma_q >0$ and $\gamma_p >0$.
The poles of the suceptibility lie at
\begin{equation}
\frac{\omega^{\rm (II)}_{\pm}}{\omega_0} \ =\ -i\frac{\gamma_q+\gamma_p}{2}\pm\sqrt{1-\left(\frac{\gamma_q-\gamma_p}{2}\right)^2} \ .
\end{equation}
It is seen that at the symmetric point $\gamma_q$ $=$ $\gamma_p$ $\equiv$ $\gamma$ the real part of the poles is just the oscillator frequency $\omega_0$. In other words, the frequency remains unchanged by the coupling to the baths. No transition to overdamped oscillations can occur.
We can compare this with the poles $\omega^{\rm (I)}_{\pm}$ for model I (see Eq.~(11) of Ref.~[\onlinecite{koh06b}]). We find the simple relation
 $\omega^{\rm (II)}_{\pm}$ $=$  $(1+\gamma_q\gamma_p)$$\omega^{\rm (I)}_{\pm}$. This indicates that the main difference between both models is a dilatation of the time scale by a factor $s =(1+\gamma_q\gamma_p)$ in model I as compared to model II.
 It is seen that when either $\gamma_q = 0$ or $\gamma_p= 0$ the poles of both models are the same.

Expression (\ref{symmetrized autocorrelationfkt}) can be compared with the corresponding one  for model I : $C_{qq}^{\rm (I)}(t,\beta)
\equiv \frac{1}{2}\langle \{ \hat{q}(t),\hat{q}(0)\}\rangle_\beta $. We find the remarkable relation
\begin{eqnarray}
C_{qq}^{\rm (II)}\left(t/s,\beta/s\right)
 &=&
s^3 C_{qq}^{\rm (I)}(t,\beta)
\ ,
\end{eqnarray}
which could have been expected from the scaling property
\begin{equation}
 \chi_{\rm II}(s\omega)\ =\ \frac{1}{s} \chi_{\rm I}(\omega) \ .
\end{equation}
This also means that the equilibrium mean square $\left<\hat{Q}^2\right>_\beta$ in model II
is enhanced by a factor $s^3$ compared with $\left<\hat{q}^2\right>_\beta$ in model I.  In the high temperature limit $C_{qq}^{\rm (I)}(t,\beta)$ becomes simple (see Eq.(22) of Ref.[\onlinecite{koh06b}]). At the symmetric point $\gamma_q=\gamma_p\equiv \gamma$ and at temperature zero ($\beta\rightarrow \infty$) it is given by
\begin{equation}
C_{qq}^{\rm (I)}(t)\ \propto \ \frac{1}{s^2} \cos\left(\omega_0 t/s\right)\exp\left(-\omega_0\gamma|t|/s\right)\ ,
\end{equation}
and therefore
\begin{equation}
C_{qq}^{\rm (II)}(t)\  \propto \  \cos\left(\omega_0 t\right)\exp\left(-\omega_0\gamma|t|\right) \ .
\end{equation}
It seems that although the Hamiltonians $\hat{H}_{\rm I}$ and $\hat{H}_{\rm II}$ are not unitarily equivalent they describe up to a time scale factor similar physics.

In summary we compared two models for a dissipative quantum system where a central oscillator couples with position and momentum to two independent heat baths. One model (model I) is physically well motivated and has found by now applications in e.g. Josephson junctions and magnetic systems. It has been studied before in detail \cite{koh05,koh06b,cuc10}. The other model (model II) has by now no obvious physical application and its study seems to be a merely academic exercise.  A more detailed study of this model will be in order if a realistic application for it can be found.

\begin{acknowledgments}
We thank F. Wegner and F. Guinea for useful discussions. This work has been supported by MICINN (Spain)
through grants FIS2007-65723 and FIS2010-21372.
HK acknowledges support from Deutsche Forschungsgemeinschaft by the
grants KO3538/1-2 and from CSIC (Spain) through Program JAE.
\end{acknowledgments}


\begin{thebibliography}{17}
\bibitem{comment1} It has been noted in Ref. \onlinecite{sanc94} that the Hamiltonian (\ref{com2a}) can be slightly modified to become fully invariant under translations.
\end{thebibliography}

\end{document}